# Hidden Risks of Unmonitored GPUs in Intelligent Transportation Systems

**Sefatun-Noor Puspa**
Clemson University
spuspa@g.clemson.edu

**Mashrur Chowdhury, Ph.D., P.E.**
Clemson University

*Abstract*—Graphics processing units (GPUs) power many intelligent transportation systems (ITS) and automated driving applications, but remain largely unmonitored for safety and security. This article highlights GPU misuse as a critical blind spot, showing how unmanaged GPU workloads silently degrade real-time performance, demonstrating the need for stronger security measures in ITS.

■INTRODUCTION Graphics Processing Units (GPUs) have become essential components in modern transportation systems, enabling intelligent infrastructure and vehicles. From smart roadside units (RSUs) equipped with video analytics to traffic surveillance cameras and advanced driver-assistance systems (ADAS), GPUs empower real-time processing of high-volume data streams. Their parallel computing capabilities support critical tasks such as object detection, license plate recognition, and traffic flow prediction. The integration of GPUs across both stationary and mobile platforms has enhanced the responsiveness in intelligent transportation applications, contributing to smarter, safer roadways and vehicles [1], [2].

This evolution represents a significant shift from traditional central processing unit (CPU)-bound systems to GPU-accelerated edge intelligence. Modern edge GPUs have been widely adopted in autonomous vehicle prototypes and edge devices [2], [3], offering substantial performance improvements for artificial intelligence (AI) workloads. Even dedicated GPUs that operate alongside the main system chip are increasingly used in research-grade automotive platforms for rapid prototyping and on-board AI inference [2]. These platforms enable complex computations at the edge, reducing latency, lowering bandwidth consumption, and increasing overall system efficiency [2].

Despite the widespread deployment of GPUs in transportation infrastructure, cybersecurity models have not kept pace with this technological shift. Traditional cybersecurity frameworks predominantly focus on protecting the operating system, CPU-bound processes, and network-level communication. As a result, GPU workloads, especially those executed directly on the GPU using programming frameworks, often fall outside the visibility of conventional endpoint detection and response (EDR) tools [4]. This blind spot creates a critical vulnerability in modern intelligent transportation systems (ITS).

GPU workloads can interfere with one another without triggering standard system alerts, allowing unmanaged or unauthorized GPU activity to run unnoticed alongside safety-critical applications. In transportation systems, such resource interference can silently drain GPU resources, leading to sustained performance degradation rather than explicit failures. The effects often appear as reduced frames per second (FPS) in vision pipelines, increased latency, or delayed responses in real-time perception tasks. Applications such as pedestrian detection or collision avoidance may continue to operate while missing timing requirements, undermining safety without being flagged by conventional monitoring tools. These silent performance failures pose direct risks to both transportation safety and infrastructure reliability.

This article addresses the gap between widespread GPU deployment and limited GPU visibility in transportation systems. It examines how unmanaged GPU activity can silently undermine real-time performance, discusses why existing monitoring approaches fail to capture such behavior, and presents a case study illustrating how performance degradation can reveal GPU misuse. The article concludes by outlining practical steps toward improving GPU security visibility in transportation platforms.

GPU IN MODERN TRANSPORTATION SYSTEMS

In ITS, GPUs play a vital role in supporting applications that require real-time, high-throughput computation. At the core of many ITS deployments is computer vision, where GPUs accelerate object detection, traffic violation monitoring, and vehicle tracking using deep learning models. Whether deployed in roadside cameras or mounted on intersection infrastructure, GPUs process large-scale visual data streams to identify pedestrians, recognize license plates, classify vehicle types, and detect red-light running or speeding incidents [1], [5]. These tasks demand rapid inference and parallel computation to meet strict latency requirements that traditional CPU-based systems cannot satisfy [2], [5]. Advanced video analytics, including motion trajectory analysis and multi-object tracking, further benefit from GPU parallelism, making GPUs indispensable for modern traffic management and situational awareness in complex roadway environments.

GPUs also support real-time sensor fusion for cooperative perception in ITS, where vehicles and roadside infrastructure share sensor data to create a more complete understanding of the surrounding traffic environment, particularly at intersections and other high-traffic areas. In these environments, multiple data sources such as light detection and ranging (LiDAR), video feeds, and vehicle-to-everything (V2X) messages must be fused to construct a comprehensive, low-latency view of the surrounding environment. Cooperative perception enables earlier hazard detection and more coordinated decision-making between smart infrastructure and connected vehicles. GPUs support this capability by executing parallel algorithms and real-time AI-based perception models. Their ability to process heterogeneous sensor data at high throughput enables safer, more effective interactions at intersections, particularly when human visibility or vehicle line of sight is limited [5].

Within vehicles, especially those supporting ADAS and autonomous driving functions, GPUs are central to perception and decision-making. Tasks such as simultaneous localization and mapping (SLAM), 3D semantic segmentation, lane detection, and trajectory planning require not only rapid computation but also energy-efficient acceleration, an area where automotive-grade GPUs excel [2], [5]. Neural networks used for obstacle avoidance, path prediction, and driver monitoring rely heavily on GPU parallelism for real-time inference. High-end autonomous platforms increasingly use neural networks for perception and decision-making. These workloads exceed the capabilities of CPUs or microcontrollers alone, making onboard GPUs essential for both safety and performance [6].

To meet the diverse computational demands of ITS applications, a range of GPU-based hardware platforms is used across the transportation ecosystem. One of those platforms is the NVIDIA Jetson Orin platform, which has become common in edge-deployed smart roadside units (RSUs) and traffic cameras, offering a compact, low-power solution with strong AI acceleration. Jetson Orin can execute full perception pipelines locally, enabling infrastructure to respond to changing traffic conditions in real time.

For autonomous vehicles, automotive-grade computing platforms such as NVIDIA's AGX platform provide an automotive-grade solution designed for mission-critical AI workloads [4]. These platforms incorporate safety-oriented features and optimized support for deep learning inference [6]. In addition, high-performance desktop GPUs are widely used in consumer desktop systems and are increasingly adopted in development kits and traffic analytics platforms. Their high computational capacity makes them well-suited for prototyping and pilot deployments before transitioning to embedded or production-grade systems.

Despite their capabilities, GPUs deployed in field-based ITS applications face distinct operational constraints. These edge devices must operate autonomously and reliably in outdoor or vehicular environments with limited connectivity, power availability, and maintenance support. For example, a roadside unit may be mounted on a traffic pole without continuous network access or physical security, making remote diagnostics and software updates challenging. As a result, GPU-based systems must not only perform their primary functions but also remain resilient to faults and adaptable to changing environmental conditions. In practice, GPUs in these settings are often underutilized relative to their computational capacity, with workloads that vary significantly based on traffic volume and time of day. This underutilization, combined with limited real-time monitoring and a lack of resource visibility, creates conditions where unmanaged GPU activity can persist unnoticed. Such activity can consume idle GPU resources, leading to overheating, sustained performance degradation, and reduced reliability of safety-critical AI workloads. These risks are frequently overlooked in traditional ITS security and system monitoring models. While GPUs greatly enhance the capabilities of transportation systems, their powerful yet low-visibility operation introduces challenges that demand greater attention from both researchers and practitioners.

GPU SECURITY RISKS

Not paying sufficient attention to GPU monitoring in ITS is no longer a hypothetical concern but an emerging challenge with direct safety and operational consequences. In GPU-accelerated platforms such as smart RSUs, traffic analytics systems, and in-vehicle computing platforms, unmanaged or unauthorized GPU activity can consume idle computing resources without the operator's awareness. Such misuse increases power draw, generates excess heat, and degrades performance, particularly during periods of low utilization. Prior studies have shown that GPU workloads can be embedded within otherwise legitimate applications and evade traditional endpoint detection mechanisms, allowing resource-intensive activity to execute unnoticed [7].

Additionally, covert computation represents a particularly concerning form of GPU misuse. Unlike CPUs, which are typically subject to process-level monitoring and host-based intrusion detection, GPU workloads often execute in isolated memory spaces using separate kernel streams [8]. This separation creates a visibility gap for many security tools. As a result, GPU kernels performing unintended or malicious computation may operate alongside benign workloads without involving the core processor or triggering conventional alerts. In machine learning environments where accelerators continuously execute complex workloads, this lack of runtime visibility increases the likelihood that unsafe or unauthorized GPU activity will go undetected [8], [9].

Another underexplored but significant risk arises from denial-of-service conditions caused by GPU resource saturation or kernel misuse. Poorly managed or intentionally abusive kernels can monopolize shared memory, saturate execution units, or block scheduling queues, leading to congestion in the GPU execution pipeline [7]. In real-time transportation systems, such conditions can have severe consequences. Vision-based systems may experience sustained frame drops or delayed detections, while autonomous driving stacks may suffer inference delays that impair perception and decision-making. These GPU-level disruptions do not always cause system crashes, but they can silently violate real-time constraints and compromise the reliability of safety-critical functions [9].

Closely related to these issues are thermal stress and long-term reliability. Unlike cloud GPUs housed in climate-controlled data centers, GPUs deployed in roadside or in-vehicle environments are exposed to temperature variation and often rely on passive or limited cooling [10]. Sustained high-load, stealthy operations can introduce continuous thermal stress, degrading hardware reliability over time and potentially triggering thermal throttling [10]. While throttling protects the device from immediate damage, it reduces computational throughput and can slow safety-critical AI workloads. In systems designed to operate for long periods with minimal maintenance, such performance degradation presents a practical reliability concern, particularly for roadside infrastructure deployed in harsh or under-maintained environments.

These vulnerabilities translate into concrete operational risks for intelligent transportation systems. When GPU resources are saturated or contended, meaning multiple workloads compete for the same processing capacity, perception tasks such as object detection from video or radar input may experience reduced frame rates or increased latency [11]. As performance degrades, critical visual cues such as pedestrians entering crosswalks or vehicles violating traffic signals may be detected too late or missed entirely. In cooperative perception systems that fuse camera, LiDAR, and other sensor data, GPU delays can lead to misaligned fusion results, producing an inaccurate representation of the surrounding environment. In connected vehicle scenarios, delays in GPU-supported processing of V2X data can disrupt time-sensitive coordination, leading to missed alerts or incorrect actions at intersections.

Finally, the operational environment of transportation edge systems further complicates risk mitigation. Many deployments rely on remote management and software updates to maintain functionality over time, while roadside devices and traffic cameras are often physically accessible. These conditions increase the likelihood that unintended or unauthorized GPU workloads may be introduced through configuration errors, software updates, or physical access, and then remain active at low intensity. When such activity does not trigger failures or alerts, its impact instead appears gradually as reduced performance, increased thermal load, and degraded reliability of safety-critical AI functions.

## WHY EXISTING MONITORING MISSES GPU MISUSE

Although GPUs are now central to real-time AI workloads in intelligent transportation systems, most monitoring and security mechanisms remain fundamentally CPU-centric. Decades of system security research have focused on observing operating systems, CPU processes, and network behavior, while GPU execution has largely been treated as a black-box acceleration layer. As a result, many forms of GPU misuse manifest only through indirect performance effects rather than explicit alerts.

Traditional approaches such as CPU hardware performance counters provide insight into instruction flow and cache behavior on the host processor, but they offer no visibility into GPU-side activity such as kernel execution, memory transfers, or resource contention. Consequently, these mechanisms cannot capture unsafe or unmanaged GPU workloads that operate independently of CPU execution paths. This gap is particularly problematic in edge-deployed transportation systems, where GPU workloads dominate perception and sensor processing.

Alternative techniques, including external power or electromagnetic sensing, have demonstrated the ability to reveal hidden computation in controlled laboratory settings. However, these approaches are impractical for real-world deployment in roadside units or in-vehicle platforms due to cost, environmental sensitivity, and integration complexity. Transportation infrastructure requires monitoring solutions that operate reliably under variable environmental conditions without additional hardware instrumentation.

While GPU telemetry interfaces, such as vendor-provided management libraries, can expose metrics on power consumption, memory usage, and utilization, access to these interfaces is often restricted or unavailable in production environments [12]. Security policies, driver limitations, and integration overhead frequently prevent continuous GPU monitoring in deployed systems. As a result, GPU behavior remains largely invisible to system operators, even as performance degradation accumulates in safety-critical applications. Also, GPU-based malware and covert computation have demonstrated how GPU resources can be abused outside traditional monitoring mechanisms [13].

Together, these limitations explain why unmanaged GPU activity can persist undetected in intelligent transportation systems. Rather than a lack of data, the core challenge lies in limited GPU monitoring and the absence of practical mechanisms to interpret performance-side effects in real time.

## CASE STUDY: INFERRING GPU MISUSE

In this case study, we examine how unmanaged GPU activity affects a real-time perception workload and whether its presence can be inferred using only application-level performance signals. The study uses a vision-based inference pipeline representative of roadside or infrastructure-deployed traffic monitoring systems. The pipeline processes video feeds collected from a roadside traffic camera installed at the South Carolina connected vehicle testbed [14] in the Clemson University campus, as shown in **Figure 1**.

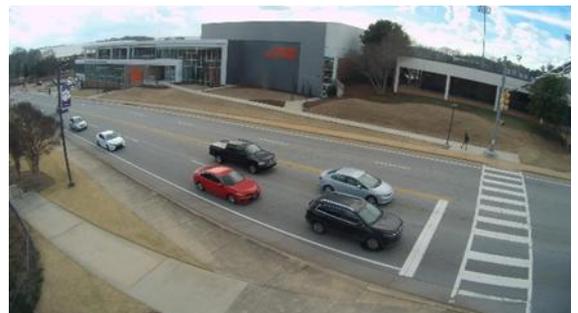

Figure 1. Example frame from the roadside traffic camera used as input to the vision-based perception pipeline.

To evaluate the effects of GPU contention, we introduce a competing GPU workload that operates independently of the perception application. This workload simulates unmanaged GPU activity that may occur in shared GPU environments where multiple processes concurrently access accelerator resources. Our objective is to observe how such contention degrades performance in the perception pipeline and to determine whether application-level signals provide meaningful insight into otherwise hidden GPU misuse under realistic deployment conditions.

**Experimental setup**

To implement the case study in a realistic transportation setting, we consider a GPU-enabled edge computing platform representative of those used in roadside AI units and infrastructure-based perception systems. The platform supports a real-time vision inference pipeline that processes continuous video streams collected from an actual roadside traffic camera and reports frames per second (FPS) as an application-level indicator of system health. The dataset consists of several 10-minute continuous traffic footage recordings at 1920x1080 resolution, reflecting common configurations used in traffic surveillance deployments.

The system operates in a shared-GPU execution environment where multiple workloads can concurrently access GPU resources. From the perspective of the vision application, GPU scheduling and resource allocation remain opaque, and no privileged instrumentation or kernel-level visibility is assumed. This reflects practical deployment conditions in which applications rely primarily on performance-side signals rather than direct GPU activity.

Table 1 summarizes the representative hardware and software configuration used to illustrate this scenario. Rather than modeling a specific attack, the setup is intended to capture the effects of unmanaged GPU activity competing with a safety-critical perception workload under realistic edge constraints.

**Table 1. Experimental Setup Summary.**

| Component | Description |
| --- | --- |
| CPU | Intel Core i7-14700K |
| GPU | NVIDIA GeForce RTX 4060 |
| RAM | 32 GB |
| Operating System | Ubuntu 22.04 LTS |
| Primary Workload | Multi-stream video-based perception |
| Competing Workload | Unmanaged background GPU workload |

For the primary workload, we implemented a multi-stream video-based perception pipeline using the YOLOv8 object detection model. The system processes multiple traffic camera video streams concurrently, performing real-time inference on incoming frames to detect objects such as vehicles and pedestrians. Running several streams simultaneously increases GPU utilization and reflects practical roadside deployments where a single edge unit may analyze feeds from multiple cameras.

To examine the effects of GPU misuse, we introduced an additional GPU workload that competes for compute and memory resources while remaining independent of the perception application. This competing workload continuously launches compute-intensive operations that occupy GPU compute units and memory bandwidth without interacting with the vision pipeline. Such behavior reflects realistic conditions in which unmanaged GPU activity can arise in shared accelerator environments, forcing safety-critical perception workloads to compete for limited GPU resources.

System behavior was evaluated under two operating conditions: baseline operation, where the perception pipeline runs in isolation, and misused operation, where unmanaged GPU activity runs concurrently with the perception workload. Performance observations focus on application-level behavior, particularly frame rate (FPS), rather than low-level GPU metrics. This approach reflects the limited visibility typically available in field-deployed transportation systems, where applications must rely on observable performance signals rather than privileged access to GPU internals.

**Results and Observations**

To evaluate the operational impact of unmanaged GPU activity, we examine several observable performance signals from the perception pipeline. These include application-level metrics such as FPS, GPU memory utilization, and system-level effects such as power and temperature changes.

**Figure 2** presents the frame rate of the perception pipeline under baseline conditions and during unmanaged GPU activity over a 500-second execution window. Under normal operation, the perception pipeline maintains stable performance at approximately 119 FPS, with only minor variability over time.

When the competing GPU workload is introduced, the frame rate drops abruptly to approximately 31 FPS. The reduction occurs immediately upon the onset of the competing activity and remains sustained throughout the contention period. Rather than fluctuating, the frame rate stabilizes at roughly one-quarter of its baseline level, indicating a substantial loss of processing capacity. Once the competing workload terminates, the frame rate returns to its original baseline level.

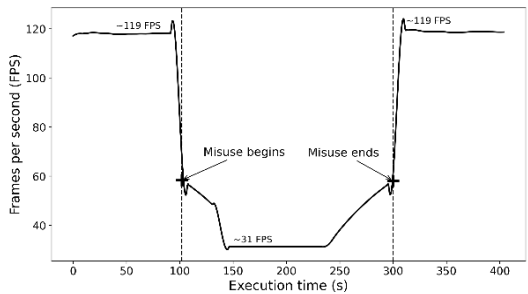

**Figure 2. Frame rate of the perception pipeline showing the reduction caused by unmanaged GPU activity.**

**Figure 3** illustrates GPU utilization during the experiment. Under baseline conditions, GPU utilization remains moderate while the perception workload processes the traffic video streams. When unmanaged GPU activity begins, utilization rises sharply and quickly saturates at nearly 100 percent. The device remains fully utilized throughout the contention period, indicating that the competing workload is occupying nearly all available compute resources. Once the competing workload stops, utilization returns to baseline levels. This sustained saturation confirms that the perception application is forced to share GPU compute resources with the unmanaged workload, which contributes directly to the observed degradation in processing performance.

**Figure 4** presents the corresponding GPU temperature measurements. During baseline operation, the device temperature begins near 34 °C and gradually increases as the perception pipeline starts processing video frames. When the competing workload becomes active, the temperature rises steadily and stabilizes around 73-74 °C, reflecting sustained high compute load on the GPU. After the competing workload terminates, the temperature decreases rapidly as the device returns to a lower utilization state. This thermal pattern further demonstrates that unmanaged GPU activity forces the device to operate under prolonged high-load conditions.

Unmanaged GPU activity not only reduces application performance and increases memory pressure but also leads to sustained increases in GPU utilization and device temperature. As shown in **Figures 3** and **4**, the competing workload drives the GPU toward full utilization and maintains the device in a prolonged high-load execution state. Sustained high utilization increases power demand and results in elevated thermal conditions. In edge-deployed systems such as roadside units and in-vehicle computing platforms, where cooling capacity is often limited, prolonged thermal stress can trigger frequency throttling mechanisms that further degrade application performance. Over time, repeated exposure to elevated temperatures may also accelerate hardware aging and reduce long-term system reliability.

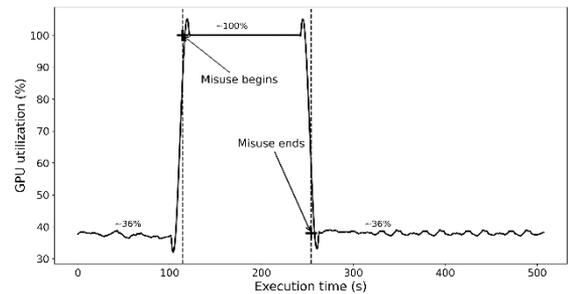

**Figure 3. GPU utilization during baseline and unmanaged GPU activity**

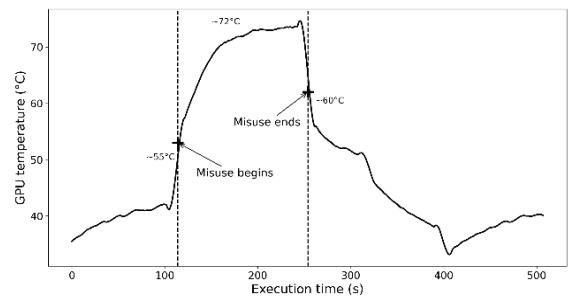

**Figure 4. GPU temperature during baseline and unmanaged GPU activity.**

**Safety and Operational Implications**

The degradation we observed in the case study represents more than a simple performance fluctuation. In real-time transportation systems, the sustained drop-in frame rate from approximately 118-120 FPS to about 31 FPS significantly increases perception latency and reduces the time available for safety-critical responses such as detecting pedestrians, vehicles, or traffic signal changes.

Importantly, the GPU remains operational throughout this period. No system crash or explicit fault occurs. Instead, the system continues running while its effective processing capacity is substantially reduced. As shown in **Figures 3** and **4**, unmanaged GPU activity drives GPU utilization to nearly 100 percent and increases device temperature from roughly 34 °C to 73-74 °C during the contention period. Increased temperature can trigger thermal stress.

This condition creates a form of silent operational degradation. Although the system appears functional from a system monitoring perspective, its ability to process sensor data in real time is significantly constrained. In transportation environments where perception accuracy and timing are critical, Such reductions in processing capacity can narrow safety margins and increase the risk of delayed or missed detections. These observations highlight the importance of monitoring GPU resource usage in intelligent transportation platforms where AI workloads depend heavily on accelerator performance.

CONCLUSIONS

This paper highlights GPU security as a critical blind spot in intelligent transportation systems. As AI-driven perception becomes central to safety-critical real-time decision-making, the GPU has evolved from a performance accelerator to a core operational component in ITS. In our case study, we demonstrated how unmanaged GPU activity can silently degrade system performance, reducing frame rates substantially while increasing memory utilization, power draw, and thermal stress, all without triggering system crashes or explicit fault signals. These effects pose serious risks to applications such as pedestrian detection, traffic monitoring, and autonomous navigation, where sustained processing performance is essential for timely response.

Our paper emphasizes that GPU misuse need not cause outright failure to introduce safety risks. Instead, it gradually degrades operational capacity, narrowing safety margins while the system appears functionally active. As ITS platforms become increasingly dependent on GPU-accelerated AI workloads, runtime visibility into GPU behavior must become a standard component of GPU monitoring and protection. Strengthening GPU observability is therefore essential not only for cybersecurity but also for maintaining functional safety and long-term reliability in connected and automated transportation systems.

RECOMMENDATIONS AND FUTURE DIRECTIONS

This study highlights the importance of treating GPU security as a critical consideration in ITS. As GPUs increasingly support safety-critical perception and decision-making, overlooking their runtime integrity creates a structural vulnerability. While current transportation cybersecurity strategies emphasize operating systems, networks, and CPUs, GPU subsystems often remain outside routine monitoring frameworks. This gap allows unmanaged or unauthorized GPU activity to degrade performance and safety margins without obvious system-level failure.

Transportation platforms should begin incorporating GPU observability into standard monitoring practices. Application-level metrics such as sustained frame rate, latency stability, and resource utilization can provide early indicators of degraded GPU operational capacity. Integrating GPU-related performance signals into existing system monitoring pipelines would offer a complete and more realistic picture of runtime health. Even without deep architectural instrumentation, continuous visibility into throughput, memory usage, power states, and thermal behavior can help identify when systems are operating outside expected performance envelopes.

Future work should extend these observations across diverse hardware platforms and deployment environments. Many transportation systems rely on embedded GPU platforms with constrained resources and limited cooling capacity. Evaluating how unmanaged GPU activity affects such systems, particularly in outdoor and vehicular conditions, is essential for understanding long-term reliability and safety implications. Analysis across different hardware platforms will help determine whether similar patterns of throughput degradation and thermal stress occur across automotive-grade and edge-oriented accelerators.

Beyond performance degradation, additional forms of GPU misuse may affect system integrity. Unintended computation, model extraction, or resource abuse could similarly introduce latency increases, memory pressure, or abnormal power behavior. Developing standardized baselines for acceptable GPU performance ranges would enable operators to distinguish normal workload variability from sustained unsafe conditions.

We further recommend that transportation safety and cybersecurity frameworks explicitly incorporate requirements for GPU monitoring. Clear thresholds for acceptable frame rate, latency, resource saturation, and thermal conditions should be defined as part of operational readiness criteria. Systems that fail to maintain real-time performance under expected workloads should be treated as operating in a degraded safety mode rather than as fully functional.

Finally, stronger collaboration between transportation operators, hardware security professionals, and hardware vendors is needed to advance GPU-aware security practices. Establishing shared benchmarks, transparent monitoring interfaces, and best practices for accelerator integrity will help ensure that GPUs are no longer treated as opaque performance enhancers but as core components of safety-critical infrastructure.

## ACKNOWLEDGMENT

This work is based upon the work supported by the National Center for Transportation Cybersecurity and Resiliency (TraCR) (a U.S. Department of Transportation National University Transportation Center) headquartered at Clemson University, Clemson, South Carolina, USA. Any opinions, findings, conclusions, and recommendations expressed in this material are those of the author(s) and do not necessarily reflect the views of TraCR, and the U.S. Government assumes no liability for the contents or use thereof.

## ■ REFERENCES

1. E. Dilek and M. Dener, "Computer vision applications in intelligent transportation systems: A survey," Sensors, vol. 23, no. 6, p. 2938, Mar. 2023, doi: 10.3390/s23062938.
2. H. Tabani, F. Mazzocchetti, P. Benedicte, J. Abella, and F. J. Cazorla, "Performance analysis and optimization opportunities for NVIDIA automotive GPUs," J. Parallel Distrib. Comput., vol. 152, pp. 21–32, 2021.
3. S. Teerapittayanon, B. McDanel and H. T. Kung, "Distributed Deep Neural Networks Over the Cloud, the Edge and End Devices," 2017 IEEE 37th International Conference on Distributed Computing Systems (ICDCS), Atlanta, GA, USA, 2017, pp. 328-339, doi: 10.1109/ICDCS.2017.226.
4. Z. Gu, E. Valdez, S. Ahmed, J. J. Stephen, L. Le, H. Jamjoom, et al., "NVIDIA GPU confidential computing demystified," arXiv:2507.02770, 2025.
5. S. Pendleton, H. Andersen, X. Du, X. Shen, M. Meghjani, Y. Eng, et al., "Perception, planning, control, and coordination for autonomous vehicles," Machines, vol. 5, no. 1, Art. no. 6, 2017.
6. Y. Miao, Y. Zhang, D. Wu, D. Zhang, G. Tan, R. Zhang, et al., "Veiled pathways: Investigating covert and side channels within GPU uncore," in Proc. 57th IEEE/ACM Int. Symp. Microarchitecture (MICRO), Austin, TX, USA, 2024, pp. 1169–1183.
7. D. Balzarotti, R. Di Pietro, and A. Villani, "The impact of GPU-assisted malware on memory forensics: A case study," Digital Investigation, vol. 14, Suppl. 1, pp. S16–S24, 2015.
8. J. Wang, Y. Wang, and N. Zhang, "Secure and timely GPU execution in cyber-physical systems," in Proc. ACM SIGSAC Conf. Comput. Commun. Security (CCS), Copenhagen, Denmark, 2023, pp. 2591–2605.
9. Y. Zhao, W. Xue, W. Chen, W. Qiang, D. Zou, and H. Jin, "Owl: Differential-based side-channel leakage detection for CUDA applications," in Proc. 54th Annu. IEEE/IFIP Int. Conf. Dependable Syst. Netw. (DSN), Brisbane, Australia, 2024, pp. 362–376.
10. M. A. Farooq, W. Shariff, and P. Corcoran, "Evaluation of thermal imaging on embedded GPU platforms for application in vehicular assistance systems," IEEE Trans. Intell. Vehicles, vol. 8, no. 2, pp. 1130–1144, 2023.
11. C. Ma, N. Wang, Q. A. Chen, and C. Shen, "SlowTrack: Increasing the latency of camera-based perception in autonomous driving using adversarial examples," arXiv:2312.09520, 2023.
12. R. Xiao, T. Li, S. Ramesh, J. Han, and J. Han, "MagTracer: Detecting GPU cryptojacking attacks via magnetic leakage signals," in Proc. 29th Annu. Int. Conf. Mobile Comput. Netw. (MobiCom), Madrid, Spain, 2023, pp. 1–15.
13. G. Vasiliadis, M. Polychronakis and S. Ioannidis, "GPU-assisted malware," in 2010 5th International Conference on Malicious and Unwanted Software (MALWARE 2010), Nancy, Lorraine, 2010, pp. 1-6, doi: 10.1109/MALWARE.2010.5665801.
14. M. Chowdhury, M. Rahman, A. Rayamajhi, S. Khan, M. Islam, Z. Khan, and J. Martin, "Lessons learned from the real-world deployment of a connected vehicle testbed," Transportation Research Record: Journal of the Transportation Research Board, 2018, doi: 10.1177/0361198118799034.

**Sefatun-Noor Puspa** is a Ph.D. student in the Glenn Department of Civil Engineering at Clemson University, Clemson, SC 29634, USA. Her research interests include cybersecurity for intelligent transportation systems, GPU security, and hardware security. Puspa received her B.S. degree in electrical and electronic engineering from the Bangladesh University of Engineering and Technology (BUET), Dhaka, Bangladesh. Contact her at spuspa@clemson.edu.

**Mashrur Chowdhury** is the Eugene Douglas Mays Chair of Transportation in the Glenn Department of Civil Engineering at Clemson University, Clemson, SC 29634, USA. He is also a Professor in the Department of Automotive Engineering and a member of the Clemson University International Center for Automotive Research (CU-ICAR). His research focuses on intelligent transportation systems, transportation cybersecurity, and connected and automated mobility. He is the Founding Director of the USDOT University Transportation Center National Center for Transportation Cybersecurity and Resiliency (TraCR) and the USDOT Center for Connected Multimodal Mobility (C2M2), and serves as Associate Director of the USDOT Center for Regional and Rural Connected Communities (CR2C2). He is also Co-Director of the Complex Systems, Analytics and Visualization Institute (CSAVI) at Clemson University.